\documentstyle[preprint,revtex]{aps}
\begin{document}
\draft
\begin{title}
Time-Dependent Current Partition in Mesoscopic Conductors\\
\end{title}
\author{M. B\"{u}ttiker}
\begin{instit}
IBM T. J. Watson Res. Ctr., P. O. Box 218,
Yorktown Heights, N. Y. 10598, and
\end{instit}
\begin{instit}
Universit\'{e} de Gen\`{e}ve,
D\'epartement de Physique Th\'{e}orique,
24 quai E. -Ansermet,
\end{instit}
\begin{instit}
CH-1211 Gen\`{e}ve, Switzerland.
\end{instit}
\receipt{\today}
\begin{abstract}
The currents at the terminals of a mesoscopic
conductor are evaluated in the presence of
slowly oscillating potentials applied to the
contacts of the sample.
The need to find a charge and current conserving
solution to this dynamic current partition
problem is emphasized.
We present results for the
electro-chemical admittance
describing the long range Coulomb
interaction in a Hartree approach.
For multiply connected samples we discuss the symmetry of the
admittance under reversal of an Aharonov-Bohm flux.
\end{abstract}
\narrowtext

\section{Introduction}

In this work we present results for the dynamic current
partition in phase-coherent,
mesoscopic conductors in the presence of
oscillating voltages applied to the contacts of the sample.
The time-dependent oscillating voltages cause
small departures of the sample away from its equilibrium state.
A specific configuration of conductors is shown in Fig. 1.
The contacts are numbered consecutively, $k= 1, 2, 3.$
The structure which connects contacts $2$ and $3$ is called
the "conductor". A second structure, which interacts with the
conductor only via long range Coulomb forces and is connected
to a single contact $k = 1,$ is called the "gate".
The currents are generated by a small
oscillating voltage $\delta V_{k} (\omega)$ applied
to the contact of the gate or one of the contacts of the conductor.
The task is to find the oscillating current
$\delta I_{k}(\omega)$ at contact $k$ of the structure.
Below we emphasize the need to find a current conserving
answer,
$\sum_{k} \delta I_{k}(\omega) = 0,$
to this problem.

For single charges moving in the electric field of an
arrangement of capacitors a current conserving answer
was given by Shockley\cite{SHOC} and Ramo\cite{RAMO}.
Clearly, in the problem of interest here, we deal not with
the motion of single charges but face a many electron problem.
Despite this, a part of the
theoretical literature takes
the discussion of Shockley and Ramo to derive
expressions for the current of a conductor
in the form of
a volume integral over the velocities of the
carriers\cite{BOOK,BULA}. Any mutual interaction between the
carriers is negelcted.
Another larger portion of the
literature is even less perceptive
and treats frequency dependent transport as a linear response
problem to an electric field which is supposed to be known.
In the most naive examples
the electric field is taken to be
uniform even for conductors which are strongly inhomogeneous
\cite{ALBER,SASO}.
The extent to which this approach is accepted is illustrated by
the fact that this portion of the literature includes authors which
are famous for their contributions to solid state physics\cite{MOTT}.

The ac-response of non-interacting
carriers is not current conserving:
Application of a time-dependent field to the sample of
Fig. 1 will give rise to currents at the terminals of the
gate and the conductor which
are unrelated to one another.
In reality currents are necessarily conserved\cite{MBAPT}.
Current conservation is closely related to the fact that
the electrical response of a conductor depends only on voltage
differences. Hence an interpretation of a linear response
calculation of independent electrons in terms of dynamic
conductance coefficients is {\it not} {\it possible}!

It is remarkable that the problem addressed here,
which is so central and basic to solid state electronics,
has found so little attention. We can cite only a few works
where the authors express a similar perception\cite{PELLE,DEVEG}.
Below we emphasize the need of a {\it charge} and
{\it current} conserving approach to ac-conductance and
exemplify this with a discussion of the ac-response of the
conductor shown in Fig. 1.
We restrict our attention to mesoscopic conductors:
It is assumed that carriers inside the sample are subject to
elastic scattering and are subject to (long range)
Coulomb forces. There is no inelastic scattering within the
sample.
The discussion of the dc conductance of mesoscopic conductors
has already taught us that
we must focus on the entire conductor including its current
and voltage contacts\cite{MB1986}.
To discuss ac-conductance we must take an even a wider view
and consider not only the conductor but also other nearby
metallic bodies\cite{MBCAP}.
In the sample of Fig. 1 the "gate" is such a nearby metallic body.
Fortunately, there is currently a growing experimental interest
in the ac-properties of small mesoscopic conductors.
A recent experiment by Pieper and Price determines both
the real and imaginary part of the ac-conductance
of a metallic diffusive Aharonov-Bohm loop
over a wide range of frequencies\cite{PPPRL}.
A theoretical discussion of these results is presented
by Pieper and Price\cite{PPPRB} using a diagramatic approach
and by Liu et al.\cite{LIUST} using the results of Ref. \cite{MBAPT}.
These results should be contrasted with earlier work
on metallic wires
by Webb et al. and De Vegvar et al.\cite{WEBB}
which found a non-linear response even at very low voltages
and very low frequencies.
For the work presented here the experiment by Chen et al.\cite{CHEN}
which measures the magnetic field symmetry of capacitance
coefficients\cite{MBCAP}
on a large two dimensional electron gas is of particular interest.
The geometry of this experiment corresponds to that of Fig. 1.
Below we state the results for the low frequency
admittance of this geometry.

The relationship between the currents
at the contacts of the sample
and the time-dependent potentials
applied to the contacts of the sample
is given by a dynamic conductance or admittance\cite{MBAPT}
\begin{eqnarray}
\delta I_{k}
= \sum_{l} G_{kl} (\omega)
\delta V_{l} (\omega).
\label{eq01}
\end{eqnarray}
We emphasize that this conductance is
a response to oscillating
electro-chemical
potentials $\mu_{k} \equiv eV_{k}$ of the contacts.
Current conservation implies that the rows and
columns of the dynamic conductance matrix
add up to zero,
\begin{eqnarray}
\sum_{l} G_{kl} (\omega)= 0,
\label{eq02}
\end{eqnarray}
\begin{eqnarray}
\sum_{k} G_{kl}(\omega)= 0.
\label{eq03}
\end{eqnarray}
The sum rules Eq. (\ref{eq02}) and (\ref{eq03})
are the same for the frequency dependent admittance as for
the elements of the dc conductance matrix\cite{MB1986}.
As in the dc-case these sum rules
of the admittance matrix
guarantee that the currents depend only on voltage
{\it differences}.

We exemplify the structure of a current conserving answer
with a discussion of the admittance to leading order
in frequency. For small frequencies we find
\begin{eqnarray}
G_{kl}(\omega)  =
G_{kl}(0)-i \omega E_{kl}
+ O(\omega^2).
\label{eq04}
\end{eqnarray}
Here
$G_{kl}(0)$ is the dc-conductance.
For the configuration of Fig. 1 the dc-conductances are related
$G_{22} = G_{33}= - G_{23}= - G_{32} \equiv
G(0)$.
All other elements of the dc-conductance matrix vanish
since the configuration of Fig. 1 allows direct transmission
only from contact $2$ to contact $3$.
In Eq. (\ref{eq04})
the term multiplying the frequency is called
an emittance\cite{MBCAP}.
In general an element of the emittance matrix might
be dominated by capacitive effects or kinetic inductive
effects resulting from direct transmission between contacts
$k$ and $l$.
The gate in Fig. 1 is coupled to the conductor
in a purely capacitive way and hence
the emittances describing this coupling are just capacitance
coefficients, $E_{kl}=C_{kl}$ if either $k=1$ or $l=1$.
There is direct transmission from contact $2$ to $3,$ and
therefore, the emittances $E_{kl}$ which connect these
contacts can in general also contain inductive effects.
Thus the emittance matrix for the conductor of Fig. 1
is of the from,
\begin{eqnarray}
{\bf E} = \left( \matrix { C_{11} & C_{12}
& C_{13} \cr
                 C_{21} & E_{22} & E_{23} \cr
                 C_{31} & E_{22} & E_{33} \cr } \right).
\label{eq05}
\end{eqnarray}
Eq. (\ref{eq02}) implies that the rows and colums of the
emittance matrix add up to zero.
Below we derive explicit expressions for the emittances and
capacitances which are current conserving.

Consider next the configuration shown in Fig. 2.
Here the "conductor" is a ring with an
Aharonov-Bohm (AB) flux threading the hole\cite{MB1983}
and connected via leads to electron reservoirs\cite{GE1984}.
The dc-conductance exhibits sample specific AB-oscillations
with fundamental period $\Phi_{0}= hc/e$.
Below we show that all coefficients of the emittance matrix
are similarly periodic functions of the AB-flux.
Even so only electrons in the "conductor" have the possibility
to circle the AB-flux long range Coulomb interactions induce
an electron density in the gate which is also a periodic
function of the AB-flux. As a consequence we predict that
the capacitance
coefficients in Eq. (\ref{eq05}) are also periodic functions
of the AB-flux\cite{MBICE}.
Of particular interest is the symmetry of
these transport coefficients under reversal of the AB-flux.
We will show that the elements of the emittance matrix
obey a reciprocity symmetry,
$E_{kl} (\Phi) = E_{lk} (-\Phi).$
This symmetry applies in particular also to the elements
of the emittance matrix which are purely capacitive.
Hence we predict that there are capacitance elements in Eq. (\ref{eq05})
which are {\it not} even functions of flux.
If we cut the conductor and consider a sample
as shown in Fig. 3, which consists of capacitors only, all elements
of the emittance matrix are purely capacitive.
Again, long ronge Coulomb interactions cause all elements of the
capacitance tensor to be periodic functions of the AB-flux.
However, in contrast to the configuration of Fig. 2, all capacitance
elements of the sample in Fig. 3 are {\it even} functions of the
AB-flux.
The key difference between the samples of Fig. 2 and 3 is that
the "conductor" in Fig. 2 represents a capacitor plate with
two contacts, whereas in Fig. 3 each capacitor plate is connected
to a single contact. Microscopically these differing symmetries
of the capacitance coefficients originate from the fact
that the capacitance coefficients in Fig. 3 represent
a long range Coulomb coupling between the full equilibrium densities
of each conductor whereas in the sample of Fig. 2 the capacitance
coefficients couple only to {\it partial} densities of the "conductor".

\section{Charge and Current Conservation}

In the presence of ac-transport current conservation
is a consequence of the long range nature of the Coulomb
interactions.
In the absence of such interactions the continuity equation
\begin{eqnarray}
-i \omega \delta n({\bf r}, \omega )
+ div \delta {\bf j}({\bf r}, \omega) =0
\label{eq09}
\end{eqnarray}
can be applied separately to the conductor and to the gate.
The volume $\Omega$ delimited by the surface shown as a broken line in
Fig. 1 defines surfaces $S_{k}$ where it intersects the reservoirs.
We can evaluate the currents across these intersections and can
evaluate the charge on the gate and on the conductor inside the
volume delimited by these surfaces.
The current at the terminal of the gate is equal to
the time derivative of the total charge on the gate
$-i\omega \delta Q_{g}(S_{1}, \omega) =
\delta I_{1} (S_{1},\omega).$
The currents at the terminals of the conductor are related to
the total charge
$\delta Q_{c}(S_{2}, S_{3}, \omega)$ on the conductor,
$-i\omega \delta Q_{c}(S_{2},S_{3}, \omega) =
\delta I_{2} (S_{2}, \omega)+
\delta I_{3} (S_{3}, \omega)$. In the absence of interactions
the currents at the conductor are not conserved nor are they
in any way related to the current at the gate.
Furthermore, the currents and charges depend on the exact
location of the surfaces $S_{k}$.
In an interacting system a time-dependent voltage applied to the
contacts of the sample leads to an internal time dependent
Coulomb potential $\delta U({\bf r}, \omega)$ and to a displacement
field $\delta {\bf d}({\bf r}, \omega)$ which is via the Poisson equation
related to the piled up charge density
$ div\delta {\bf d}({\bf r}, \omega) = -4 \pi
\delta n({\bf r}, \omega )$.
If the Poisson equation is combined with the continuity equation,
Eq. (\ref{eq09}),
then the total current density is conserved,
\begin{eqnarray}
div \delta {\bf j}({\bf r}, \omega ) = 0.
\label{eq10}
\end{eqnarray}
Here the total current density is the sum
of the "displacement" current and the "particle" current,
$\delta {\bf j}({\bf r}, \omega ) = - i(\omega/4\pi)
\delta {\bf d}({\bf r}, \omega ) + \delta j_{p}({\bf r}, \omega )$.
The total current density has no sinks or sources.
Consider the volume $\Omega$ as shown in Fig. 1.
The total current flowing across the surface $S$
into this volume is zero.
For an arbitrary choice of the volume $\Omega$
the total current density is not spatially limited to the contacts of the
conductors.
But assume that the contacts are good metallic
conductors which screen any electric fields over distances
of a Thomas-Fermi screening length. This permits us to chose
the volume $\Omega$ so large that no electric field lines
will penetrate its surface. We can assume that
the surface of $\Omega$ intercepts the electron reservoirs
at a distance which is sufficiently far from the connection
of the contact to a lead.
At such distances electric fields are screened efficiently.
Now if there are no electric field lines penetrating the surface
of $\Omega,$ the current must be confined to the reservoirs.
Current conservation now means that the sum of all currents
in the contacts of the sample must add up to zero,
\begin{eqnarray}
\sum_{k}
\delta I_{k} (\omega) = 0.
\label{eq11}
\end{eqnarray}
Since there are no electric field lines which penetrate
the surface of $\Omega$ the total charge in $\Omega,$
according to Gauss,
is also zero,
\begin{eqnarray}
\delta Q(\omega) = 0.
\label{eq12}
\end{eqnarray}
For the conductors of Fig. 1, Eq. (\ref{eq12}) implies that
any charge accumulation on the conductor is compensated
by an equal charge of opposite sign on the gate.
If the capacitance of the conductor vanishes, then the total
charge on the conductor must vanish.
In that case, application of an ac-voltage can polarize
the conductor but cannot lead to a total fluctuating charge
on the conductor.
In the non-interacting case, the currents
and charges depend on the exact location of the surfaces,
irresepctively on how far out in the reservoir the surfaces
are located. In contrast,
in the interacting case, the currents are independent
of the exact choice of $\Omega,$ once the intersections are
deep in the reservoirs.

The task at hand is not trivial since the electric
field or potential must be known to find the correct answer.
The potential is, however, not a single particle property, but
depends on the location of all mobile carriers, their
separation among themselves and their separation
to the ionic lattice, sheets of donors etc.
The solution of this problem depends, therefore, on the
approach taken to handle the many body problem.
We presented solutions using three different
simple schemes:
(a) A discrete potential model approximates the potential
landscape with the help of a finite number of
discrete potentials.
In the simplest case only one potential
per conductor is introduced\cite{MBAPT}.
(b) For metallic conductors with efficient screening the
local potential landscape
is determined in Thomas-Fermi screening
approximation\cite{MBHTP}.
(c) For conductors as shown in Fig. 1 for which
long range Coulomb forces matter, the potential landscape
is determined in a Hartree-like approximation\cite{MBCAP,MBICE}.
This list should eventually be extended to a
Hartree-Fock approach
to include exchange and to density functional
theory to include exchange and correlations.

Quite generally we approach the task as follows\cite{MBAPT}:
First we derive an external response: The currents are found
in response to an oscillating {\it chemical} potential at a contact.
During this step the Coulomb interaction is switched off but
the response is calculated with the help of the wave functions
of the self-consistent equilibrium problem.
In a second step the Coulomb potential caused by the additional
injected charges is evaluated. In a third step the current
response due this internal potential is found.

\section{External response due to oscillating chemical potentials}

In electrical conductors one has to distinguish electro-chemical
potentials $\mu_{k},$ chemical potentials $E_{Fk}({\bf r})$
and the electro-static potential $eU({\bf r})$.
At equilibrium these potentials are related,
$\mu_{k} = E_{Fk}({\bf r}) +
eU({\bf r})$.
In a first step we evaluate the response to an electro-chemical
potential oscillation
$edV_{k}(t) = d\mu_{k} (t)$
assuming that it is caused by an oscillation in the
chemical potential
$edV_{k}(t) = d\mu_{k} (t)
= dE_{Fk}(t)$
in contact $k$.
In reality, of course, the contact, except where it narrows into
a lead, has to be charge neutral.
To achieve this we will then in a second step calculate the
potential and restore charge neutrality in the contacts.
To obtain the leading order frequency dependent terms in the
external response we
proceed as described below.

We assume that the conductors of interest are described by
scattering matrices ${\bf s}_{kl}$ which give the  out going
current amplitudes
in the asymptotic states\cite{MB1991}
of contact $k$ in terms the incoming
current amplitudes in contact $l$.
For the conductor of Fig. 1, the scattering matrices are,
\begin{eqnarray}
{\bf S} = \left( \matrix { {\bf s}_{11} & 0 & 0 \cr
                 0 & {\bf s}_{22} & {\bf s}_{23} \cr
                 0 & {\bf s}_{32} & {\bf s}_{33} \cr } \right).
\label{eq13}
\end{eqnarray}
The matrix ${\bf S}$ is unitary and due to
micro-reversibility\cite{MB1991} we have
${\bf S}^{\star}(-\Phi) = {\bf S}^{-1} (\Phi)$ and consequently
${\bf S}^{T}(\Phi) = {\bf S}(-\Phi).$
The elements of the scattering matrix are explicit
functions of the energy
of the incident carriers, the
AB-flux (except for the elements of ${\bf s}_{11}$),
and are a functional of the electrostatic potential $U(\bf {r}).$
(The self-consistent potential $U$ also depends on the AB-flux and
depends on the electro-chemical potentials).
To brief we will not write s-matrix elements as a function or
functional of all these quantities but only emphasize the
argument most important for the immediate discussion.
The density of states at the Fermi energy in the gate is
\begin{eqnarray}
\frac{dN_{11}}{dE} =
\frac{1}{4\pi i}\,
{\rm Tr} \left[ {\bf s}^{\dagger}_{11} (E)
\frac{d{\bf s}_{11} (E)}{dE}
-\frac{d{\bf s}^{\dagger}_{11} (E)}{dE}
{\bf s}_{11}(E)\right]
\label{eq14}.
\end{eqnarray}
The density of states in the conductor is a sum of four
contributions,
\begin{eqnarray}
\frac{dN_{kl}} {dE} =
\frac{1}{4\pi i}\,
{\rm Tr} \left[ {\bf s}^{\dagger}_{kl} (E)
\frac{d{\bf s}_{kl} (E)}{dE}
-\frac{d{\bf s}^{\dagger}_{kl} (E)}{dE}
{\bf s}_{kl}(E)\right]
\label{eq15}
\end{eqnarray}
where $k,l = 2, 3$.
The scattering matrix $s_{kl}$ determines
the out going current
amplitudes in contact $k$ as a function of the current amplitudes
of the incident waves. Thus
the density of states given by Eq. (\ref{eq15})
represent a {\it partition} of the total density of states
according to the origin (injecting contact) and destination
(emitting contact) of the sample.
(A portion of the total density of states of the conductor
might consist of localized states which are disconnected from
any contact. At this stage the localized states do not play a role.
Later on, to determine the screening properties of the sample,
we will have to consider not only the mobile states, but also
the localized states).

A small increment in the Fermi energy
$dE_{F1}(t)$
at the gate injects an additional charge
$dQ(t)=(e dN_{11}/{dE})dE_{F1}(t)$ onto the gate.
This charge flows to the gate through contact $1$ and
gives rise to a current at this contact given by
$dI_{1}(t) = dQ/dt$. For a sinusoidal variation
of the chemical potential
$dE_{F1}(t) = dE_{F1} exp(-i\omega t)$
we find thus at contact $1$ a current
\begin{eqnarray}
dI_{1}(\omega) = -i \omega e^{2} (dN_{11}/{dE}) dV_{1}(\omega),
\label{eq16}
\end{eqnarray}
where
$edV_{1}(\omega) =
dE_{F1}(\omega)$
has been used.
Consider now the conductor with two contacts.
A chemical potential variation
$edV_{2}(\omega) =
dE_{F2}(\omega)$
at contact $2$ leads to an
additional charge
$(e dN_{32}/{dE}+
e dN_{22}/{dE})
edV_{2}(\omega)$
on this conductor.
Now the fact that we have scattering matrix expressions
(see Eq. (\ref{eq15})) for
these charges helps us further.
Apparently it is only the additional charge
$dQ_{3}(\omega) = e (dN_{32}/{dE})
edV_{2}(\omega)$ which leads to a current at contact 3,
whereas
$dQ_{2}(\omega) = e (dN_{22}/{dE})
edV_{2}(\omega)$ is determined by carriers which leave
the conductor through contact 2.
Therefore, charging of this conductor causes currents
\begin{eqnarray}
dI_{k}(\omega) = -i \omega e^{2} (dN_{kl}/{dE}) dV_{l}(\omega)
\label{eq17}.
\end{eqnarray}
at its contacts.
Since direct transmission between contact $2$ and $3$
is possible, an oscillating voltage causes in addition
at these contacts a current determined by the dc-conductance.
Thus the leading low frequency-current response
to an oscillating chemical potential
$edV_{k}(\omega)$
is given by an external response\cite{MBAPT}
\begin{eqnarray}
G^{e}_{kl}(\omega) = G_{kl}(0)
-i \omega e^{2} (dN_{kl}/{dE}).
\label{eq18}
\end{eqnarray}
This external response is not current conserving.
Since for the dc-conductances we have
$\sum_{k} G_{kl}(0)=
\sum_{l} G_{kl}(0)=0,$
we find that to leading order in frequency
$\sum_{k} G_{kl}^{e}(\omega)$
is proportional to the total charge injected from contact $l$
into the conductor. This charge is determined by the
{\it injectance}
$\sum_{k}(dN_{kl}/{dE})$ of contact $l$.
Similarly we find that to leading order in frequency
$\sum_{l} G_{kl}^{e}(\omega)$
is proportional to the total charge that is emitted by the
conductor through contact $k.$
This charge is determined by the (unscreened)
{\it emittance}
$\sum_{l}(dN_{lk}/{dE})$ of contact $k$.
Only if
for some reason the injectances and emittances vanish
would we have current conservation.
We mention only in passing that in the absence of a magnetic
field the injectance and emittance of each contact is equal.

The injected charges create an internal, time-dependent
Coulomb potential
$\delta U({\bf r}, t).$ This time-dependent electro-static
potential will in turn cause additional currents.
We next investigate the response to such an internal
electro-static potential.

\section{Response to an oscillating electro-static potential}

We are interested in the currents generated in the contacts
of a sample in the presence of an oscillating
potential
$\delta U({\bf r}, t).$
We can Fourier transform this potential
with respect to time
and consider a perturbation of the form
$u({\bf r}) (U_{+\omega} exp(-i\omega t) +
U_{-\omega} exp(+i\omega t))$. Since the potential is real
we have $U_{-\omega}=U^{\ast}_{+\omega}$.
The response to such a potential can be treated using
a scattering approach\cite{MBHTP}: Due to the oscillating
internal potential a carrier incident with energy $E$
can gain or loose modulation energy $\hbar \omega$
during reflection at the sample or during transmission
through the sample.
The amplitude of an out going wave is
a superposition of carriers incident at energy $E$ and
at the side-band energies,
$E \pm \hbar \omega.$
In the low-frequency limit of interest here the
amplitude of the out going waves can be obtained
by considering the scattering matrix
${\bf s}_{kl}
(U({\bf r}, t), E)$
to be a slowly varying function of the potential
$U({\bf r}, t).$
Since the deviations of the actual potential away from the
(time-independent) equilibrium potential
$U_{eq}({\bf r})$ are small, we can expand
the scattering matrix in powers of
$\delta U({\bf r}, t) =
U({\bf r}, t) -
U_{eq}({\bf r}).$
To linear order the amplitudes of the out going waves
are determined by
${\bf s}_{kl}
(U({\bf r}, t), E) =
{\bf s}_{kl}
(U_{eq}({\bf r}), E) +
(\delta {\bf s}_{kl}/
\delta U({\bf r}))
\delta U({\bf r}, t)$.
Evaluation of the current at contact $k$ gives\cite{MBHTP}
\begin{eqnarray}
\delta I_{k}(\omega) =  i e^{2}
\omega
\int d^{3}{\bf r} (dn(k, {\bf r})/dE)
u({\bf r})
U_{+\omega}.
\label{eq19}
\end{eqnarray}
Here we have introduced the (local) density of states
\begin{eqnarray}
dn(k ,{\bf r})/dE =
- (\frac{1}{4 \pi i}) \sum_{l}
Tr\left[{\bf s}^{\dagger}_{kl}
(\frac{\delta {\bf s}_{kl}}{\delta eU({\bf r})})
- (\frac{\delta {\bf s}^{\dagger}_{kl}}{\delta eU({\bf r})})
{\bf s}_{kl}(E) \right]
\label{eq20}
\end{eqnarray}
of carriers at point ${\bf r}$ which are emitted by the
conductor at probe $k.$
We refer to this density of states as {\it emissivity}
of the sample into contact $k$.
In Eq. (\ref{eq20}) the summation over $l$ invokes a single
term $l=1$ for the gate and invokes
the terms $l=2$ and $l=3$ for the currents at the contacts
of the conductor.
The scattering matrices are evaluated at the
equilibrium chemical potentials $E = \mu_{1}$ of the gate
and $E = \mu _{2} = \mu_{3}$ of the conductor.
For later reference we also introduce the {\it injectivity}
\begin{eqnarray}
dn({\bf r},l)/dE =
- (\frac{1}{4 \pi i}) \sum_{k}
Tr\left[{\bf s}^{\dagger}_{kl}
(\frac{\delta {\bf s}_{kl}}{\delta eU({\bf r})})
- (\frac{\delta {\bf s}^{\dagger}_{kl}}{\delta eU({\bf r})})
{\bf s}_{kl}\right]
\label{eq21}
\end{eqnarray}
where again the summation invokes just $k=1$ for the gate and
invokes both $k = 2$ and $k=3$ for the conductor.
A more detailed
derivation of Eq. (\ref{eq19}) can be found in
Ref. \cite{MBHTP}.
It is useful to express the response to the internal potential
in the from of a conductance defined as
$\delta I_{k}(\omega) =
G^{i}_{k}(\omega)
U_{+\omega}$.
Comparison with
Eq. (\ref{eq20}) gives for the internal conductances
\begin{eqnarray}
G^{i}_{k}(\omega) = i e^{2}
\omega
\int d^{3}{\bf r} (dn(k, {\bf r})/dE)
u({\bf r}).
\label{eq22}
\end{eqnarray}
We note that the internal conductance, in contrast  to the
external conductance, contains no dc-contribution.
A static potential that vanishes far out in the leads
cannot produce a dc-current\cite{KANE,NOECKEL}.
A static potential variation in the interior
of a conductor just brings us from a conductor with one equilibrium
potential configuration to another conductor with another
equilibrium potential configuration.
The absence of a dc-term
in Eq. (\ref{eq22}) re-emphasizes
that the dc-transport is a consequence of differences
in electro-chemical potentials at the contacts rather
than an acceleration of carriers due to an electric field.

In the presence of an oscillating voltage at contact $l$
the current at contact $k$
has two contributions determined by the external response
$G^{e}_{kl}$ and by the internal response
$G^{i}_{k}.$ But the internal response given by
Eq. (\ref{eq22}) contains an as yet undetermined internal potential.
To complete the calculation of the total
response we must find the dependence of the internal potential
on the oscillating electro-chemical potentials of the contacts.
Fortunately, in the low-frequency limit of interest here,
it is sufficient to find the dependence of the internal potential
on the electro-chemical potentials in quasi-static approximation.

\section{Characteristic Potentials}

Let us first consider a steady state of the configuration of Fig. 1.
We allow for dc-transport and will consider the time-dependence later.
The electro-static potential
$U(\mu_{1},\mu_{2},\mu_{3}, {\bf r})$
for these conductors is
a function of the electro-chemical potentials,
and a complicated function of position.
Small increases in the electro-chemical potentials
$d\mu_{1},
d\mu_{2},
d\mu_{3}$
will bring the conductor to a new state with
an electro-static potential
$U(\mu_{1}+d\mu_1,\mu_{2}+d\mu_2, \mu_{3}+d\mu_{3})$.
The difference $dU$ between these two potentials
can be expanded in powers of the increment in electro-chemical potential.
To linear order we have
\begin{eqnarray}
edU(\mu_{1}, \mu_{2},\mu_{3}, {\bf r}) =
\sum_{k} u_{k}({\bf r}) d\mu_{k}.
\label{eq23}
\end{eqnarray}
Here
$u_{k}({\bf r}) =$
$edU(\mu_{1}, \mu_{2}, \mu_{3}, {\bf r})/
d\mu_{k}|_{d\mu_{k} = 0},$ with $k = 1, 2$
are the {\it characteristic}
{\it potentials}\cite{MBCAP}.
These characteristic functions determine the electro-static potential
inside the sample in response to a variation of an electro-chemical
potential at a contact.
These static characteristic potentials are what we need to
complete our calculation.

The characteristic potentials are the solutions of a Poisson
equation with a non-local screening kernel:
\begin{eqnarray}
-\Delta u_{k} ({\bf r}) + 4 \pi e^{2}
\int d^{3}r^{\prime}
\Pi ({\bf r}, {\bf r}^{\prime})
u_{k} ({\bf r}^{\prime}) =
4 \pi e^{2} (dn_{k}({\bf r})/dE)_{U}.
\label{eq24}
\end{eqnarray}
The source term
$(dn_{k}({\bf r})/dE)_{U}$ is the {\it injectivity} of contact $k$:
The charge density injected into the conductor as a consequence of
a small increment in the electro-chemical potential at contact $k$.
The injectivity is given by Eq. (\ref{eq21}).
The screening kernel is determined by the charge density response
function $\Pi$, which to be brief, we call Lindhard function.
A potential acting on the sample,
in the absence of long range
Coulomb interactions, generates an induced charge density
\begin{eqnarray}
dn_{ind} ({\bf r}) = - \int d^{3}r^{\prime}
\Pi({\bf r}, {\bf r}^{\prime}) e dU({\bf r}^{\prime}).
\label{eq25}
\end{eqnarray}
Note that the Lindhard function describes the variation of the
charge density not only of the mobile electrons which can be
reached from the contacts but also of the localized states
which might exist inside the conductors or the gate or somewhere
in the volume $\Omega$.
If the source term
$e(dn_{k}({\bf r})/dE)_{U}$
in Eq. (\ref{eq24}) is replaced
by a test charge
$ e\delta({\bf r}-{\bf r}_{0})$
which is concentrated at one point ${\bf r}_{0},$
the solution to Eq. (\ref{eq24}) is
Green's function $g({\bf r}, {\bf r}_{0}).$
With the help of Green's function
we find for the characteristic function,
\begin{eqnarray}
u_{k} ({\bf r}) =
\int d^{3}r^{\prime} g ({\bf r}, {\bf r}^{\prime})
(dn_{k}({\bf r}^{\prime})/dE)_{U}.
\label{eq26}
\end{eqnarray}
Since an equal increment of the electro-chemical potential at
all contacts by $d\mu$ only changes the energy scale but must leave
all densities unchanged\cite{MBAPT}, the charcteristic functions
must add up to one at every point\cite{MBCAP},
\begin{eqnarray}
\sum_{k} u_{k} ({\bf r}) = 1.
\label{eq27}
\end{eqnarray}
This implies for Green's function,
the property\cite{MBCAP}
\begin{eqnarray}
\int d^{3}r^{\prime} g ({\bf r}, {\bf r}^{\prime})
\sum_{k} (dn_{k}({\bf r}^{\prime})/dE)_{U} = 1.
\label{eq28}
\end{eqnarray}
The same relationship follows
from the condition that the sum of all induced
charge densities
plus the test charge is zero.
Eq. (\ref{eq28}) is an important result of this section:
It will be used to demonstrate
charge and current conservation of the admittance.

\section{The Electro-Chemical Emittance Matrix}

The charcateristic potentials determine the quasi-static
potential inside the conductor.
In response to an oscillating voltage
$edV_{k}(\omega) exp(-i\omega t) = d\mu_{k} (\omega) exp(-i\omega t)$
the internal potential is given by
$edU(\mu_{1}, \mu_{2},\mu_{3}, \omega, {\bf r}) =
\sum_{k} u_{k}({\bf r}) d\mu_{k}(\omega)$.
Inserting this into the expression for the internal conductance
we find upon comparison with Eq. (\ref{eq09})
the following expression for the emittance:
\begin{eqnarray}
E_{kl} =
e^{2}
(dN_{kl}/dE)
- e^{2}
\int d^{3}r
\int d^{3}r^{\prime}
(dn(k, {\bf r})/dE)
g ({\bf r}, {\bf r}^{\prime})
(dn({\bf r}^{\prime},l)/dE).
\label{eq29}
\end{eqnarray}
To understand Eq. (\ref{eq29}) it is useful to consider for
a moment the configuration in Fig. 3. Here we have cut the
"conductor". Thus in this configuration there is no transmission.
The emittance coefficients given by Eq. (\ref{eq29}) determine
the electro-chemical
capacitance matrix of this configuration,
$E_{kl} = C_{kl}.$
In this configuration the first term in Eq. (\ref{eq29}) is
non-vanishing only in the diagonal terms.
The non-diagonal terms are negative. Furthermore, since
each capacitor is coupled to a single contact the
injectivity and emissivity in Eq. (\ref{eq29}) are identical,
$dn(k, {\bf r})/dE =
dn({\bf r},k)/dE.$
Hence the capacitance matrix has the symmetry $C_{kl} = C_{lk}$.

For the conductors shown in Fig. 1 and 2 the topmost row
and the first column of the emittance matrix consists of puerly
capacitive elements since the first term in Eq. (\ref{eq29})
vanishes.
However the emittance elements
$E_{22}, E_{32}, E_{23}$ and $E_{33}$
represent a competition between two effects:
The first term in
Eq. (\ref{eq29}) is a kinematic term due to the fact
that we have direct transmission from contact $l$ to $k$.
It is a kinetic inductance which is counterbalanced
by the second, capacitive term in
Eq. (\ref{eq29}).
Depending on the transmission properties of the sample and the
screening properties of the conductor either the kinetic inductance
can dominate and give a positive emittance or the capacitive effects
can dominate and the non-diagonal elements of the emittance are negative.

\section{Magnetic-Flux Symmetry of the Emittance Matrix}

In this section we discuss briefly the symmetry of the
low frequency transport coefficients under reversal of the
magnetic flux.
The electro-chemical emittances
are periodic functions of the AB-flux $\Phi$
with fundamental period $hc/e$.
Here we are concerned with the symmetry of these coefficients
as the polarity of the flux is reversed.
First consider the geometry of Fig. 3 in which all conductors
are coupled purely capcitively.
In this case the emissivities and injectivities in each conductor
are identical,
$dn(k, {\bf r})/dE =
dn({\bf r},k)/dE$ and represent a change in the
{\it equilibrium} density of the conductor. But the equilibrium
density of the conductor is an even function of flux,
$dn(\Phi, {\bf r})/dE =
dn(-\Phi, {\bf r})/dE$.
The same conclusion follows immediately from the scattering
matrix expressions for these densities.
Similarly, since the Linhard function gives the change in
density as we go from one equilibrium state to another, the
Lindhard function is also an even function of flux.
Consequently, in the purely capacitive geometry of Fig. 3 the
electro-chemical capacitance coefficients are all
even functions of the flux,
$C_{kl} (\Phi) = C_{kl}(-\Phi)$.
We have shown earlier that the capacitance coefficients in this
configuration are also symmetric,
$C_{kl} (\Phi) = C_{lk}(\Phi)$.

Consider next the conductor of Fig. 2. The Lindhard function
is again an even function of flux.
But now the emissivities and injectivities of the
"conductor" are not identical. Instead they obey
a reciprocity relationship,
$dn(\Phi,{\bf r},k)/dE =$
$dn(-\Phi, k, {\bf r})/dE.$
This is seen most directly by using the scattering matrix expressions
Eqs. (\ref{eq18}) and (\ref{eq21})
for these densities.
As a consequence the electro-chemical emittance matrix has the
reciporcity symmetry,
$E_{kl} (\Phi) = E_{lk}(-\Phi)$.
Only the diagonal elements of the emittance matrix are even functions
of the flux. Of particular interest is the fact that
this symmetry also
applies to the purely capcitive elements of the emittance
matrix. Thus for the conductor in Fig. 2 the
capacitance elements obey
$C_{kl} (\Phi) = C_{kl}(-\Phi)$.
There are thus purely capacitive elements of the emittance matrix
of the conductor in Fig. 2
which are {\it not} even functions of magnetic flux.
This is in stark contrast to the capacitance matrix of the conductor
in Fig. 2.

In the geometry of Fig. 3, the injected charge is determined
by a single density of states
$dn_{k}(\Phi,{\bf r})/dE$ which is equal to the injectance
and is equal to the emittance of this conductor.
In contrast in the conductor of Fig. 2 the capacitance
coefficients and the emittances are determined by the
injectivities and emittances of differing contacts.
Each of these densities represents only a portion of the
total local density of states (of mobile carriers).
The total local density in the conductor is equal to
the sum of the injectivities of both contacts,
$dn({\bf r},2)/dE +
dn({\bf r},3)/dE $
and is equal to the sum of the emissivities into both contacts
$dn(2, {\bf r})/dE+ dn(3, {\bf r})/dE.$
The capacitance elements in the geometry of Fig. 3 are
even functions of flux since the total local density of states
counts. In contrast the differing properties of the capacitance
and emittance elements of the conductor of Fig. 2 are due to the
fact that the emissivities and injectivities represent  only a portion
of the total local density of states.

This novel prediction of the symmetry of capacitance elements
was tested in an experiment by Chen et al. \cite{CHEN}.
The symmetry predictions for a conductor in a uniform magnetic field
are the same as those discussed above for an AB-flux.
The conductor (a high mobility two-dimensional electron gas) used by
Chen et al. has the same topology
as the conductor in Fig. 1.
A gate was made which only partially overlaps the edge of the conductor.
The capacitance elements of the gate with respect to the contacts
of the two dimensional electron gas are shown to exhibit
a very dramatic asymmetry: The full geometrical capacitance is
obtained for one polarity of the field whereas the capacitance
elements nearly vanish for the other field polarity.
The emittance elements which describe dynamic corrections
to the dc-conductance
are harder to measure than the capacitance coefficients.

\section{Discussion}

We have emphasized the need for a current conserving
approach to frequency dependent conductances. An illustration
of such a theory for low frequencies has led to novel
transport coefficients which are determined by effective, local
density of states.
Measurements of these transport coefficients provides information
on the charge and potential distribution inside the sample.

Clearly, experiments which demonstrate an AB-effect in
an electro-chemical capacitance
would be fascinating. Such experiments would highlight
that these coefficients should not be treated as geometrical
constants, but like conductances, represent information on
the physical properties of samples.
An electro-chemical emittance has been
measured in the experiment by Pieper and Price\cite{PPPRL}
and has been shown to exhibit AB-oscillations over the entire
accessible range of frequencies\cite{LIUST}.
AB-oscillations in a capacitance coefficient are likely to
be an even more challenging task since the AB effect in the capacitance
is likely only a small correction to a geometrical capacitance.

\figure{A gate at electro-chemical potential $\mu_{1}$
coupled capacitively to a conductor with contacts at
electro-chemical potentials $\mu_{2}$ and $\mu_{3}$.
A Gauss volume $\Omega$ is chosen such that no electric field
lines penetrate its surface.
\label{autonum}}

\figure{A gate coupled to a doubly connected conductor.
An AB-flux $\Phi$ threads the hole of the loop.
As a consequence of the long range Coulomb interaction
the densities of the conductor and of the gate depend
on the AB-flux.
\label{autonum}}

\figure{Arrangement of capacitors each connected to a single
electron reservoir. One capacitor plate is a loop threaded
by an AB-flux $\Phi$.
All elements of the electro-chemical capacitance
matrix are (even) functions of flux.
\label{autonum}}

\end{document}